\documentclass[namedreferences,12pt]{kluwer}
\begin{document}
\begin{article}
\begin{opening}
\title{The Magnetic Field Sign Reversal and Evolution of Rossby and Alfv\'en Waves
Induced by Velocity Shear Oscillations on the Sun}
\author{A.D. \surname {Pataraya}}
\institute{Department of Theoretical Astrophysics, Abastumani Astrophysical Observatory,\\
$2^a$, Kazbegi Ave., Tbilisi 380060, Georgia}
\author{T.A. \surname{Pataraya}}
\institute{Department of Physics, Tbilisi State University, Chavchavadze ave. 2, 380028 Tbilisi, Georgia}
\author{B.M. \surname{Shergelashvili}}
\institute{Department of Theoretical Astrophysics, Abastumani Astrophysical Observatory,\\
$2^a$, Kazbegi Ave., Tbilisi 380060, Georgia\\ 
The Abdus Salam International Centre for Theoretical Physics, Strada Costiera 11, 34100 Trieste, Italy}
\runningauthor{A. D. Pataraya, T. A. Pataraya and B. M. Shergelashvili}
\runningtitle{The Magnetic Field Sign Reversal and Evolution of Rossby and Alfv\'en Waves in the Solar Interior}
\begin{abstract}
The behaviour of the toroidal and meridional components of the solar
large-scale magnetic field and linear Alfv\'en and Rossby waves during
solar activity cycles and bi-annual time periods are theoretically investigated
in this work. The shear of the toroidal and meridional components of velocity
is taken into account. We consider the case of periodical velocity shear
with bi-annual oscillation period, hereinafter the Velocity Shear Quasi Bi-annual
Oscillations (VSQBO). The large-scale magnetic field toroidal
and meridional components are obtained as harmonic functions of the time
and the oscillation phase difference between them is equal to $\pi/2$.
The sign reversal of the magnetic field toroidal and meridional components
is studied. The numerical simulations show that in the case of significant VSQBO amplitude
values the toroidal or meridional component of the large-scale
magnetic field reverses its sign three times in one of the hemispheres
(northern or southern) of the Sun, during the solar activity
cycle 23.  According to our results the appearance of velocity shear oscillations
leads to the modulation of the magnetic field 22-year period oscillations by the
bi-annual ones. The presented model is applicable for investigation of the magnetic
field evolution at the base of convection zone as well as for understanding the
magnetic field properties in the upper solar atmosphere (the magnetic canopy of the
solar chromosphere \inlinecite{Ro96}).

The excitement of the linear Rossby and Alfv\'en waves in the shear layer at the base of
the convection zone is also considered. The periodical impulsive growth is characteristic
to the energy density of Alfv\'en and Rossby waves and they propagate as localized in time
powerful pulses. Properties of these pulses are governed by the direction of waves phase velocity and the
initial phase of VSQBO, and such behavior of waves well explains mechanisms of the solar
flare excitement and activity.
\end{abstract}
\keywords{Magnetic Field, Waves, Activity Cycles, Flare Activity, Sun} 
\end{opening}
\newpage
\section{Introduction}
The large-scale magnetic field longitudinal and latitudinal distribution and sign
reversal during the solar cycle  23 have been investigated by many authors (\opencite{to20};
\opencite{bek99}; \opencite{be96}; \opencite{be92}; \opencite{bai95}; \opencite{bu89};
\opencite{ga83}; \opencite{vi82}; \opencite{am73}). The recent observations of the p-mode frequency
splitting and results of corresponding helioseimic inversions of the solar internal rotation show
the appearance of two different layers with strongly marked shear flow (e.g. \opencite{ko97};
\opencite{se98}). One of these layers is located at  the base of convection zone and the differential
rotation in this region, as is well known, plays an important role in the solar dynamo mechanism.
The second one  is located just below the photosphere. It is also reasonable to
think that layer effects on the distribution of the global magnetic field at the solar
surface and  takes place in the formation of canopy of magnetic field in the upper solar
atmosphere \inlinecite{Ro96}. In  other words the differential rotation takes an imposing  part  in
the formation and evolution of the large-scale solar magnetic field. 

    In this work we present the ideal magnetohydrodynamic (MHD) model explaining three-time sign reversal
of the large-scale magnetic field at the beginning of cycle 23. It is known from the theory of
the solar dynamo \inlinecite{pa79}, that  the toroidal  and meridional components of the large-scale magnetic
field oscillate with constant phase difference of $\pi/2$: the toroidal magnetic field  reaches its maximum value
when the meridional one is minimum or vice versa. In our model we obtain toroidal and meridional
magnetic field components as harmonic functions of the time, when the differential
rotation and its time oscillation are taken into account. Hereinafter in the
work, the periodic time variation of the velocity shear with 2-year period is denoted as the Velocity Shear Quasi Bi-annual
Oscillations (VSQBO). We present the result of numerical simulations showing that our
model can explain one-time and three-time sign reversals of the large-scale magnetic field components
during the activity cycle 23 and, in general, the modulation of activity cycle 23 by oscillations
with a period of about 2 years. The behaviour of the large-scale magnetic field strongly depends on
the amplitude and initial phase of VSQBO. The similar model  was developed by \inlinecite{pa97}
using  VSQBO. In that case the toroidal component of the magnetic field was a linear
function of time and the meridional one was a constant. The structure of the large-scale magnetic field
cyclic oscillations is also considered in the several works  (\opencite{be94}, \citeyear{be95},
\citeyear{be98}; \opencite{ak87}). 

    We analytically investigate the possible mechanisms of the formation and evolution of the active regions and
the corresponding excitement of the solar flares. In this sense, excitement
of solar flares and their periodical activity are well known from several observations (e.g. see
\opencite{bai87}; \opencite{ri84}). In this work we investigate the propagation of the linear
Rossby and Alfv\'en waves in the environment with the steady and periodically oscillating velocity
shear. Additionally, toroidal and meridional components of the magnetic field
change sign one or three times during the activity cycle 23. We consider the time evolution
of these waves. The numerical simulations show that the hydrodynamic and the magnetic
energy densities oscillate with short periods and the waves propagate as localized in time pulses.
These results are in agreement with the observed periodicity of solar flare activity and this
allows one to think that our model is applicable for the explanation of solar flares excitement and activity.    
\section{Basic Equations}
In this work we investigate the behaviour of the large-scale solar magnetic field in the layer,
with the differential rotation and excitement of linear waves in this layer. As it is evident
from helioseismic observations there are two kinds of such layers in the solar interior. One of them 
is at the base of the convection zone and the other is just below the photosphere. These layers
represent the areas bounded by two concentric spherical surfaces. The calculations are
done in the co-rotating Cartesian frame. $xoy$ plane is tangential to
the bottom boundary of the layer. The $x$
axes is directed eastward and it represents the toroidal direction, the $y$ axes is
directed northward and it represents the meridional direction and the $z$ axes is directed 
outward to the upper solar atmosphere and it determines the radial
direction. We use the ideal MHD equations for incompressible medium \cite{pr82}. So, for given
latitude we have a set of differential equations in the following form:
\begin{equation}
\frac {DB_x}{Dt} =B_x\frac {\partial V_x}{\partial x} + B_y\frac {\partial V_x}{\partial y}, 
\label{f1}
\end{equation}
\begin{equation}
\frac {\partial B_x}{\partial x} + \frac {\partial B_y}{\partial y} = 0,
\end{equation}
\begin{equation}
\frac {DV_x}{Dx} = -\frac {1}{\rho} \frac {\partial p}{\partial x} + 
\frac {B_y}{4\pi \rho} \biggl ( \frac {\partial B_x}{\partial y} - \frac {\partial B_y}{\partial x}\biggr ) 
+(f_0 + \beta y)V_y,
\end{equation}
\begin{equation}
\frac {DV_y}{Dt} = -\frac {1}{\rho} \frac {\partial p}{\partial y}
- \frac {B_x}{4\pi \rho}\biggl ( \frac {\partial B_x}{\partial y} -
\frac {\partial B_y}{\partial x} \biggr ) - (f_0 +\beta y)V_x,
\end{equation}
\begin{equation}
\frac {\partial V_x}{\partial x} +\frac {\partial V_y}{\partial y}=0,
\label{f5}
\end{equation}
where,
\begin{equation}
\frac {D}{Dt} = \frac {\partial}{\partial t} +
V_x \frac {\partial}{\partial x} +V_y\frac {\partial}{\partial y}.
\end{equation} 
In these equations  $V_x$, $B_x$ are the toroidal and $V_y$, $B_y$ are the meridional components
of the velocity and the magnetic field, respectively; $\rho = {\rm const}$ and $p$ are density and pressure;
$f_0 = 2\Omega \sin \theta$ is Coriolis parameter;
\begin{equation}
\beta = \frac {2}{R_ \odot} \frac {\partial}{\partial \theta}(\Omega \sin \theta),
\label{f7}
\end{equation}
where, $\Omega$ is the angular rotation frequency of the sun;  $R_ \odot$ is the Solar radius and $\theta$
is the latitude corresponding to the origin of the local coordinate frame. The positive values of $\theta$
correspond to the northern hemisphere and the negative ones denote latitudes in the  southern hemisphere.
The other symbols have their usual meanings.

It is well known from the solar models that the rotation frequency of the Sun is a function of latitude
$\Omega = \Omega (\theta)$ and it could be well approximated as (\opencite{br89}; \opencite{ho70}):
\begin{equation}
\Omega = \Omega _0 (1-A_1 \sin ^2 \theta -A_2 \sin ^4 \theta),
\end{equation}
where, $\Omega_0$, $A_1$ and $A_2$ are constants. It should be noted that by using 
eq. (\ref{f7}) we obtain magnitude of the latitude, where $\beta$ is equal to zero. After substituting
$A_1$ and $A_2$ values we have $\beta =0$ at the following latitudes:
 \begin{equation}
 \theta ^{(N)} _1 \approx 74^{\circ} 33 ^{'},\hskip2cm
 \theta ^{(S)} _1  \approx -74^{\circ} 33 ^{'}
 \end{equation}
As we have mentioned above our aim is to study the excitement of the linear Rossby and Alfv\'en
waves in the layer where steady velocity shear and VSQBO appear. But before we turn to
this matter let us determine the mathematical form of quantities satisfying the (\ref{f1})-(\ref{f5}) set of
equations in an unperturbed state of environment. 
\section{The Sign Reversal of the Magnetic Field Components}
    We consider the differential rotation of the Sun with respect to the co-rotating Cartesian frame
of reference described above. In this case  we have the toroidal component of the unperturbed
velocity in the following form:
\begin{equation}
V_{x0} = yV_{xy},
\label{f10}
\end{equation}
where, $V_{xy}$ is the shear in the $y$ direction of the toroidal flow and it has the form:
\begin{equation}
V_{xy}=V+V_{xn}\cos (\omega _n t+\varphi _0)=Vf_2,
\label{f11}
\end{equation}
The first term on the right side of eq. (\ref{f11}), $V$ is the permanent shear of the toroidal flow in the considered layer and it 
conveys the steady differential rotation of the Sun: 
\begin{equation}
V = \frac {d}{d\theta}(\Omega \cos \theta) =-\Omega _0 \sin \theta (1+2A_1+(4A_2-3A_1)\sin ^2 \theta -5A_2\sin ^4 \theta).
\label{f12}
\end{equation}
This expression was presented in \inlinecite{pz95} and \inlinecite{gi95}. [In these works
the excitement of linear waves were investigated, when the toroidal component of the unperturbed
magnetic field is a linear function of time.] 

By using eq. (\ref{f12}) we can obtain the latitude value, 
where the steady velocity shear of the toroidal flow $V$ has a maximum by solving equation
$dV/d\theta=0$, and after substituting of $A_1$ and $A_2$ values we obtain:
\begin{equation}
\theta ^{(N)} _2 \approx 55^{\circ} 9^{'}, \hskip2cm
\theta ^{(S)} _2  \approx -55^{\circ} 9^{'}
\label{f13}
\end{equation}

The other term on the right side of eq. (\ref{f11}) determines the velocity shear oscillation
with an arbitrary period and in the case of 2 year oscillation period it demonstrates the VSQBO.
In this term $\omega _n$ is the frequency of the velocity shear oscillations:  
\begin{equation}
\omega _n =\frac {2\pi}{nt_0}, 
\label{f14}
\end{equation}
where $t_0\approx 3.16\cdot 10^7$sec$=1$ year. $nt_0$ is the oscillation period of the velocity shear.

We have the meridional component of the unperturbed velocity in the form:
\begin{equation}
V_{y0} = xV_{yx},
\label{f15}
\end{equation}
where $V_{yx}$ is the shear of the meridional flow in the $x$ direction: 
\begin{equation}
V_{yx}= -\frac{\Omega^2 _1}{V}f_2,
\label{f16}
\end{equation}

According to eqs. (\ref{f10})-(\ref{f12}), (\ref{f15})-(\ref{f16}) and the equation of induction we
obtain expressions for the toroidal $B_{x0}$ and meridional $B_{y0}$ components of the
unperturbed magnetic field:
 \begin{equation}
B_{x0} = B_{xm}\sin (\Omega _1 t_0 f_1),
\label{f17}
\end{equation}
\begin{equation}
B_{y0} = B_{ym}\cos (\Omega _1 t_0 f_1),
\label{f18}
\end{equation}
where, $\Omega _1= 2\pi/22t_0$ is the frequency of low-frequency oscillations (with the period
22 years) of the unperturbed toroidal and meridional components of the solar
magnetic field; $B_{xm}=B_0\sin \theta = V\cdot B_{ym}/\Omega _1$ here, $B_0={\rm const}$
\begin{equation}
f_1 = \tau -1 + a_n [\sin (\frac {2\pi}{n} \tau +\varphi_0)-\sin (\varphi_0)],
\end{equation}
\begin{equation}
f_2 = 1 + \frac {V_{xn}}{V} \cos(\frac {2\pi}{n} \tau +\varphi_0),
\end{equation}
where $\tau = t/t_0$ and 
\begin{equation}
a_n = \frac {V_{xn}}{\omega _n t_0  V} .
\label{f21}
\end{equation}
As we can see from eqs. (\ref{f17}) and (\ref{f18}) the toroidal and meridional components
of the magnetic field are harmonic functions of the time and they oscillate with a constant phase difference
of $\pi/2$. We carried out the numerical simulations for the case of $n=2$ in eq. (14). In other words,
we consider the VSQBO. We studied the time evolution of the magnetic field components defining the VSQBO amplitude in
two different ways:

a) Using in eqs. (\ref{f11}) and (\ref{f21}) the following expression of the $V_{xn}$:
\begin{equation}
V_{xn}=V_0\sin^2 \theta \cos \theta,
\label{f22}
\end{equation}
 one has 
\begin{equation}
a_2 = -\frac {V_0}{\pi \Omega _0} \frac {\sin \theta \cos \theta}{1+2A_1 +(4A_2-3A_1)\sin ^2 \theta -5A_2 \sin^4 \theta}.
\label{f23}
\end{equation}
Analysing  the results of numerical simulations we found that the low-frequency (with 22-year period)
oscillation of the large-scale magnetic field components is modulated by the high-frequency (with bi-annual
period) one. The modulation rate depends on the latitude $\theta$ and the magnitude of $a_2$. Particularly,
the significant modulation of the magnetic field oscillation takes place, when $\pi \vert a_2 \vert >1$ and 
the three-time sign reversal of the magnetic field toroidal component occurs at the beginning of the activity cycle.
Figure 1 shows the time evolution of the normalized toroidal $B_{x0}/B_{xm}$ (solid lines) and the meridional
$B_{y0}/B_{ym}$ (dashed lines) components. Figure 1a corresponds
to the latitude value, where the steady velocity shear of the toroidal flow has maximum $\theta =\theta ^{(N)} _2$
in the northern hemisphere and Figure 1b corresponds to the same latitude in the southern hemisphere
$\theta =\theta ^{(S)} _2$ (see eq. (\ref{f13})). In this case we assume that the amplitude of the VSQBO
$V_{xn}$ is defined by eq. (\ref{f22}) and its initial phase $\varphi_0=\pi$. These results show
that at the middle latitudes the amplitude of VSQBO is large enough to induce the significant modulation of the
large-scale magnetic field cyclic oscillations by the bi-annual ones in both hemispheres of the Sun. This modulation
leads to the magnetic field toroidal component three-time sign reversal within the first two years of the activity
cycle 23 in the northern hemisphere (see the encircled area 1 in Figure 1a), while in the southern one 
only one-time sign reversal occurs during the same time interval. As one can see from the presented Figures, when
$\varphi_0=\pi$ at the solar activity minimum the three-time sign reversal of the magnetic field toroidal component
appears only in the northern hemisphere. We have a similar picture in the southern hemisphere in the time
interval between 11 and 13 years from the beginning of the cycle 23. The time behaviour of the magnetic field
toroidal component is inversed within this time interval, i.e. now three-time sign reversal occurs only
in the southern hemisphere (see the encircled area 1 in Figure 1b). In other words, the modulation of the magnetic
field components oscillation is not synchronized between different hemispheres and, because of this,
we have the asymmerty in the three-time sign reversal occurrence. This kind of behaviour is characteristic of
the magnetic field toroidal component for most latitudes except the areas very close to the equatorial
plane and pole. We will address this in more detail later.
With regard to the magnetic field meridional component (the dashed line curves), its low-frequency
oscillation is also strongly modulated by the bi-annual one in both hemispheres of the Sun. In spite of this,
the three-time sign reversal of the meridional field component  disappears for most of the latitude
values, although this effect can almost be achieved only close to the latitudes where the steady velocity
shear is maximum (eq. (\ref{f13})) and this effect occurs in the northern hemisphere in the time interval
$6<\tau <8$ years (see the encircled area 2 in Figure 1a) and in the southern one in the time interval
$5<\tau <7$ years (see the encircled area 2 in Figure 1b). It is important to note that,
if we perform calculations for the case of $\varphi_0=0$, then we obtain an inversed picture, i.e.,
in this case the three-time sign reversal of the magnetic field toroidal component occurs only in the southern
hemisphere, within the first two years of activity cycle.

As we mentioned above the situation changes in the equatorial and polar areas. The time evolution
of the magnetic field components in these areas is very similar in both hemispheres, we therefore present
the numerical results regarding the northern hemisphere. In Figures 1c and 1d the curves
for latitudes close to the equatorial plane ($\theta = 5^\circ$) and to the pole ($\theta = 85^\circ$) in the northern
hemisphere are presented, respectively. The behaviour of the large-scale magnetic field in the equatorial and polar
areas is different from that considered above. In these areas the amplitude of the VSQBO is small and, because of this,
$\pi \vert a_2 \vert$ becomes $<1$. As one can see in Figures 1c and 1d, in this case the
very weak modulation of the magnetic field low-frequency oscillation occurs and the above mentioned sign reversal
does not appear. Finally, exactly on the equatorial plane ($\theta = 0^{\circ}$) or on the poles
($\theta = \pm 90^{\circ}$) the modulation of the cyclic oscillation vanishes completely. 

b) We have also performed the numerical simulations for the case when the amplitude of
the VSQBO $V_{xn}$ is defined as:
\begin{equation}
V_{xn}= -U_0V \sin \theta (\delta (1-(\omega_n t_0)^{-1})+\sin \theta),
\label{f24}
\end{equation}
and in this case we obtain $a_2$ as follows:
\begin{equation}
a_2= -\frac {U_0\sin \theta (\delta (1-\pi^{-1})+\sin \theta)}{\pi},
\label{f25}
\end{equation}
where $U_0$ and $\delta$ are constants. As in the first case, we present in Figure 2
the curves showing the time dependence of the normalized toroidal and meridional magnetic field
components, when the VSQBO amplitude is defined by eq. (\ref{f24}), its initial
phase is $\varphi_0=\pi$, $\delta=1.4$ and $U_0=1.8$. Figure 2a corresponds to the latitude
value in the northern hemisphere, $\theta =\theta ^{(N)} _2$, and Figure 2b to that in the southern
one, $\theta =\theta ^{(S)} _2$. These results show, that unlike the previous case, when
$\delta=1.4$ the strongly marked modulation of the magnetic field component low-frequency oscillation
appears (i.e. $\pi \vert a_2 \vert>1$) only in the northern hemisphere, while in the southern one
$\pi \vert a_2 \vert<1$ and the modulation is weak. Hence, the three-time sign reversal
of the magnetic field toroidal component within first two years of the activity cycle occurs only in the
northern hemisphere (see the encircled area 1 in Figure 2a), while in the southern hemisphere the modulation is
very weak and thus the three-time sign reversal is absent in the whole hemisphere. Unlike the previous case, the asymmetry in
the magnetic field toroidal component three-time sign reversal is not due to different phases of modulation of
the low-frequency oscillations, but now the strong modulation occurs only in the northern hemisphere.
As it is shown in Figure 2a the oscillation of the meridional component is modulated too, but in spite of this
its three-time sign reversal does not occur at most of latitude values. It is important to note that,
we can obtain an inversed picture in the case, when $\delta =-1.4$ and $\varphi _0=0$. i.e. in this case 
we have the strong  modulation of the magnetic field cyclic oscillation only in the southern hemisphere. 

Finally, the behaviour of the magnetic field components close to the equatorial plane is similar to that
shown in the previous case. As one can see from Figure 2c, the above mentioned modulation is very weak
and thus the sign reversal effect does not occur near the equatorial plane. Although, from eq. (\ref{f25})
one can see that $\vert a_2 \vert$ increases with latitude $\theta$ and when $\delta >0$ it is maximum on the northern pole
and when $\delta <0$ it is maximum on the southern pole. Hence, in these cases the three-time sign reversal the magnetic
field toroidal component appears at one of the polar areas too (see the encircled area 1 in Figure 2d). Moreover, in this case
the three-time sign reversal of the meridional component can also be achieved (see the encircled area 2 in Figure 2d).
\section{The Evolution of Rossby and Alfv\'en Waves}
In the previous section we studied (\ref{f1})-(\ref{f5}) a set of differential equations in the unperturbed state
and we determined the expressions for the toroidal and meridional components of the 
unperturbed flow velocity and magnetic field. Let us now consider the excitement
of the linear waves in the incompressible medium located in the solar internal shear layer.
In the unperturbed state we take into account the shear and VSQBO (eqs. (\ref{f10}) and (\ref{f15})).
Additionally, the components of the magnetic field are oscillating with period of 22 years. At the same time this
low-frequency oscillation  is modulated by the high-frequency oscillation of the magnetic field  with period equal
to that of VSQBO.

We investigated equations (\ref{f1})-(\ref{f5}) by using the linear perturbation theory. So, all quantities are
presented in the form $\psi =\psi _0 +\psi ^{'}$ and we obtain the following set of the linearized
differential equations : 
\begin{equation}
\frac {\partial B^{'} _x}{\partial x}+\frac{\partial B^{'} _y}{\partial y}= 0,
\label{f26}
\end{equation}
\begin{equation}
\frac {\partial V^{'} _x}{\partial x}+\frac{\partial V^{'} _y}{\partial y}= 0,
\end{equation}
\begin{equation}
\frac {D_0 B^{'} _x}{D_0 t}=\Biggl (B_{x 0}\frac {\partial}{\partial x}+B_{y 0}\frac {\partial}{\partial y}\Biggr )V^{'} _y +V_{yx}B^{'} _x,
\end{equation}
\begin{equation}
\frac {D_0 }{D_0 t}\Biggl ( \frac {\partial V^{'} _x}{\partial y}-\frac {\partial V^{'} _y}{\partial x}\Biggr ) =
\frac {1}{4\pi \rho}\Biggl (B_{x 0}\frac {\partial}{\partial x}+B_{y 0}\frac {\partial}{\partial y}\Biggr )
\Biggl ( \frac {\partial B^{'} _x}{\partial y}-\frac{\partial B^{'} _y}{\partial x}\Biggr ) +\beta V^{'} _y,
\label{f29}
\end{equation}
where,
\begin{equation}
\frac {D_0 }{D_0 t}= \frac {\partial}{\partial t}+V_{x0}\frac {\partial}{\partial x}+
V_{y0}\frac {\partial}{\partial y}.
\end{equation}
As we can see from eqs. (\ref{f10}) -(\ref{f15}) both $V_{x0}$ and $V_{y0}$ depend on the time
and spatial coordinates. On the other hand the magnetic field components are harmonic functions of time (eqs. \ref{f17}-\ref{f18}).
Because of this, in the equations we have temporal as well as the spatial inhomogeneity. In this case 
the modal analysis is not applicable for the investigation of the perturbed quantities. 

We studied eqs. (\ref{f26})-(\ref{f29}) by using the non-modal perturbation theory which is considered by many
authors (\opencite{ch93}; \opencite{lo88}; \opencite{fa87}; \opencite{mp77}; \opencite{ty31}). According to
non-modal analysis, when we consider the linear velocity shear profile, it is possible to find
out the coordinate transformation, which in the linearized equations, leads to the changing of spatial inhomogeneity
by a temporal one. Using this coordinate transformation allows one to investigate the temporal evolution of the
spatial Fourier harmonics and in this case all of the perturbed functions have the following form: 
\begin{equation}
\psi ^ {'}= \psi _1 exp \{ i K_x x + iK_y y \},
\label{f31}
\end{equation}
where, 
\begin{equation}
K_x = k_{x1}F_1,
\label{f32}
\end{equation}
\begin{equation}
K_y = \frac {V}{\Omega_1}k_{x1} F_2,
\label{f33}
\end{equation}
are the toroidal and meridional components of the wavevector, respectively. In expressions (\ref{f32}) and (\ref{f33}):
\begin{equation}
F_1 =C K_{21}\sin(\Omega _1 t_0 f_1) + \cos(\Omega_1 t_0 f_1)
\label{f34}
\end{equation}
\begin{equation}
F_2 = -\sin(\Omega t_0 f_1) + C K_{21}\cos(\Omega t_0 f_1),
\label{f35}
\end{equation}
where, $C=\Omega_1/V$ and $K_{21}=k_{y1}/k_{x1}$ 
In these equations $k_{x1}$ and $k_{y1}$ are constants. According to
expressions (\ref{f26})-(\ref{f29}) and (\ref{f31})- (\ref{f33}) we obtain the set of equations for
perturbed functions in the following form:
\begin{equation}
K_xB_{x1} + K_y B_{y1}=0,
\label{f36}
\end{equation}
\begin{equation}
K_xV_{x1} + K_y V_{y1}=0,
\label{f37}
\end{equation}
\begin{equation}
\frac {dB_{y1}}{dt} = i\sqrt {4\pi \rho}\omega _A V_{y1}-\frac {K_y}{K_x} V_{yx}B_{y1},
\label{f38}
\end{equation}
\begin{equation}
\frac {d}{dt} \Bigl ( \frac {K^2 _{*}}{K_x}V_{y1}\Bigr )= i \beta V_{y1}+\frac {i\omega _A}{\sqrt {4\pi \rho}}
\frac {K^2 _{*}}{K_x}B_{y1},
\label{f39}
\end{equation}
where,
\begin{equation}
K^2 _{*}=K^2_x+K^2 _y
\end{equation}
and $\omega _A$ is the frequency of Alfv\'en waves and it has the following form:
\begin{equation}
\omega _A =\frac {K_x B_{x0}+K_yB_{y0}}{\sqrt {4\pi \rho}} = \frac {B_{xm}}{\sqrt {4 \pi \rho}} C k_{y1},
\end{equation}
where, $B_{xm}$ is the amplitude of the magnetic field toroidal component (see eq. (\ref{f17})).
First, we consider the case when $k_{y1}=0$. In this case $\omega _A =0$ and propagation of only Rossby waves
is possible and we obtain the analytical solution $V_{y1}=V ^{(R)}_{y1}$ of eq. (\ref{f39}) in the following form:
\begin{equation}
V^{(R)}_{y1}=V_{y1}(0) \frac {\cos (\Omega_1 t_0 f_1)}{1+(C^{-2} -1)\sin ^2(\Omega_1 t_0 f_1)} \exp \Bigl \{-\frac {i \beta}{k_{x1}(C^{-2}-1)^{1/2}}arctg [(^{-2}-1)^{1/2}\sin (\Omega_1 t_0 f_1)] \Bigr \}
\label{f42}
\end{equation}
Since all perturbed quantities are assumed to be $\sim \exp (i\varphi)$ and the oscillation frequency 
$\omega=-\partial \varphi/\partial t$ we obtain the dispersion relation of Rossby waves:
\begin{equation}
\omega_R = -\frac {i\beta \cos (\Omega_1 t_0 f_1)}{k_{x1}(1+(C^{-2} -1)\sin ^2 (\Omega_1 t_0 f_1))}
+xk_{x1}\Omega_1 f_2\sin (\Omega_1 t_0 f_1) +yk_{x1}\Omega_1 f_2\cos (\Omega_1 t_0 f_1),
\end{equation}
and at last, by using of eqs. (\ref{f37})and (\ref{f42}) we have the expression of Rossby waves hydrodynamic energy
density $W_R =\rho(\vert V^{(R)} _{x1}\vert ^2 + \vert V^{(R)} _{y1} \vert ^2)/2$ in the following form:
\begin{equation}
W_R = \frac {\rho V^2 _{y1}(0)}{2} \frac {1}{1+(C^{-2}-1)\sin ^2 (\Omega_1 t_0 f_1)}
\end{equation}
In Figure 3 the time dependence of the $W^R$ normalized by its initial value is presented. Figures 3a and 3b
correspond to the northern ($\theta =\theta ^{(N)} _2$) and southern ($\theta =\theta ^{(S)} _2$) hemispheres,
respectively, when the amplitude of the VSQBO is defined by eq. (\ref{f22}) and its initial phase $\varphi _0 =\pi$.
As we can see from these figures, when in the medium propagate only Rossby waves their hydrodynamic
energy periodically increases in short time period during the solar activity cycle, which is demonstrated by peaks in the plots.
The time moments of the
wave energy density peak appearance coincide with those when the unperturbed magnetic field toroidal component 
changes its sign. As we have mentioned above, in this case occurs three-time sign reversal of the toroidal component and
this fact reveals itself in the three peaks at the beginning of the activity cycle 23 (Figure 3a). The peaks
denoted by "1" in Figures 3a and 3b represent the the moment of sign reversal due to the mean low-frequency
oscillations of the magnetic field toroidal component and the occurrence of peaks denoted by "2" and "3" is caused
by the three-time sign reversal effect. Hence, when one-time sign reversal occurs one has only the single peaks. 
In the moments of sign reversal of the unperturbed magnetic field toroidal component the Rossby waves
propagate as pulses localized in time. [It should be noted that by comparison of Figure 3a and 3b it is clearly
seen that the asymmetry between the northern and southern hemispheres due to the fact that modulation of the magnetic
field cyclic oscillation in the different hemispheres is asynchronous.] 

  We have also performed the numerical simulations in the case when the amplitude $V_{xn}$ of the VSQBO is
defined by eq. (\ref{f24}), its initial phase $\varphi_0=\pi$, $\delta =1.4$ and $U_0=1.8$.
Resulted curves presented in Figure 4 showing the time dependence of the Rossby wave hydrodynamic
energy density normalized by its initial value. Figures 4a and 4b correspond to the northern ($\theta =\theta ^{(S)} _2$) and
southern ($\theta =\theta ^{(S)} _2$) hemispheres, respectively. In these Figures we
have the peaks of energy density in the time moments when sign reversal of the unperturbed magnetic field toroidal
component occurs. As we know from the above discussion in this case we have three-time sign reversal only in the northern
hemisphere and this leads to the appearance of two additional peaks denoted by "2" and "3" beside the mean peak denoted by "1"
in Figure 4a. Furthermore, we have only single peaks in the southern hemisphere and this is a result of the disappearance of the
strong modulation of the magnetic field cyclic oscillations by bi-annual ones. This property of waves determines the
asymmetry of solar flare activity intensity within the solar hemispheres and a detailed discussion on this matter
is presented below. 

It is important to note that, in general, in the case of $k_{y1}=0$, the meridional component of the Rossby wave
wavevector is $K_y=0$ at the time moments when the sign reversal of the unperturbed
magnetic field toroidal component occurs.
This means that, in time moments when the Rossby waves propagate exactly in the toroidal direction the
wave energy density increases and it has the form of a sharp peak. 

Now let us consider the case when $k_{y1} \neq 0$. In this case, in the medium, propagation of
both Rossby and Alfv\'en waves is possible. We performed a numerical calculation of the (\ref{f36})-(\ref{f39}) set of 
differential equations. In Figure 5 the normalized hydrodynamic and magnetic energy densities are presented as
functions of time corresponding to $\theta = \theta ^{(N)} _2$ (Figures 5a) and $\theta = \theta ^{(S)} _2$ 
(Figures 5b) values of the latitude. One can see from these curves that
wave hydrodynamic
\begin{equation}
W_H = \frac{\rho}{2}(\vert V_{x1}\vert^2 +\vert V_{y1}\vert^2)
\end{equation}
and magnetic
\begin{equation}
W_M = \frac {\vert B_{x1}\vert^2+ \vert B_{y1}\vert^2}{8\pi}
\end{equation}
energy densities periodically increase during the short time intervals while in the initial momentum we assumed
that the hydrodynamic and magnetic energy densities are commensurable $W_H (0)/W_M (0) \approx 1$.
The linear  Alfv\'en waves propagate as powerful pulses. The period of energy density oscillation
depends on the $k_{x1}$ and $k_{y1}$ magnitude. In other words the hydrodynamic 
and magnetic wave energy densities oscillate with period which depends on the direction of the wave phase
velocity in the $\tau=0$ time moment. Figure 5 shows curves for $k_{x1}=10^{-9}$ cm$^{-1}$
and $K_{21}=2$.  The calculations are carried out for values of environment parameters satisfying those obtained from
the standard solar models for the base of the solar convection zone. Particularly, the amplitude value of the
unperturbed magnetic field toroidal component (eq. (\ref{f17})) is determined assuming $B_0=10^5$ G. 
Figure 6 shows the time variation of the hydrodynamic and the magnetic energy densities for the
same values of parameters, but in this case the amplitude of the VSQBO is taken by eq. (\ref{f24}).
As it is clear from the curves in Figure 5 and Figure 6 the character of the wave temporal evolution is
similar in both cases. This model well explains the mechanism of the solar flares excitement and activity.
For the different values of $k_{x1}$ and $k_{y1}$ the
wave energy density oscillation with periods well describing the observed 51-day (e.g. see \opencite{bai87})
as well as 154-day (e.g. see \opencite{ri84}) periodicity in the occurrence of the solar flare activity may be obtained.
Observational data  \cite{bai87} also show asymmetry in intensity of the flare activity between northern and southern
hemispheres of the Sun. Comparing curves demonstrated in Figures 5a and 5b (Figures 6a and 6b)
one can see that the amplitude of energy density oscillation in northern hemisphere is greater than that in southern hemisphere
and the more intensive excitement of the solar flares in the northern hemisphere is expected. According to
our results the flare activity is more intensive in the hemisphere where the thee-time sign reversal of the
unperturbed magnetic field toroidal component occurs within the first two years of the activity cycle.  

As we have shown in section 2 in the unperturbed state the modulation of large-scale magnetic field
low-frequency oscillation  vanishes close to the polar and equatorial areas, when the VSQBO amplitude is
determined by eq. (\ref{f22}). This leads to the decrease of the wave energy oscillation amplitude in mentioned
areas. On the other hand, when we consider the case of the VSQBO amplitude by eq. (\ref{f24})
then in the polar area of the northern hemisphere the amplitude of the wave energy density oscillations remains
to be significant.
 
   So, we have presented the model of solar flare excitement at the base of the solar convection zone 
as the reasonable mechanism of the solar flare activity origin. As we mentioned in the introduction our
model is also applicable for the shear layer just below the photosphere. It is reasonable to think that the
velocity shear effects on the large-scale magnetic field distribution at the solar surface and the excitement
and  the time evolution of Rossby and Alfv\'en waves take part in the formation of the magnetic field structure
in the upper solar atmosphere and in the heating of the solar corona.
\section{Conclusions}
  In this work we investigated the behaviour of the large-scale magnetic field in the solar internal shear layers.
The VSQBO is taken into account. By use of the numerical simulations we studied the behaviour of the
magnetic field toroidal and meridional components. The excitement and time evolution of the linear Rossby and Alfv\'en waves
at the base of convection zone were also studied, when the VSQBO and modulation of the magnetic field components
were taken into consideration. We obtained results for our analytical solutions and for the above mentioned numerical simulations,
which can be formulated as follows:

1) We obtained the expressions of the magnetic field toroidal and meridional components
in the form of  harmonical functions of time.

2)   We obtained the curves representing the time dependence of the magnetic field toroidal and meridional
components in two cases:

a) When the VSQBO  amplitude is defined by eq. (\ref{f22}). In this case, as shown in Figures 1a and 1b,
we have the strongly marked modulation of the large-scale magnetic field low-frequency (with period 22-years) oscillation
by the high-frequency (with the 2-year period) ones at most latitudes of both northern and southern
hemispheres of the Sun. When the initial phase of VSQBO $\varphi_0=\pi$, the modulation
of the magnetic field components cyclic oscillation leads to three-time sign reversal of the toroidal component
in the northern hemisphere, within the first two years after activity minimum. During the same time interval we
have only one-time sign reversal of the magnetic field toroidal component in the southern hemisphere. This asymmetry,
in occurrence of the three-time sign reversal effect between different hemispheres, is caused by asynchronous
modulation phase in these hemispheres. The three-time sign reversal of the magnetic field meridional component does
not happen at most latitude values. Exceptions are areas where the steady shear of the velocity toroidal
component is maximal (see eq. (\ref{f13})). In this case the three-time sign reversal of the meridional component
may happen in time interval between 6 and 8 years after activity minimum in the northern hemisphere and
between 5 and 7 years in the southern hemisphere.

 The above mentioned effect of the modulation decreases in equatorial and polar areas (Figures 1c and 1d).
 Because of this, in these areas three-time sign reversal of any magnetic field component becomes impossible.
 
 b) We also considered another case, when the VSQBO amplitude  is defined by eq. (\ref{f24}).
Corresponding curves are presented in Figure 2. In this case, when the initial phase of VSQBO
$\varphi_0=\pi$ the strongly marked modulation of the magnetic field components low-frequency
oscillation appears only in the northern hemisphere (see Figure 2a), while in the southern hemisphere this modulation
is very weak (Figure 2b). Because of this the three-time sign reversal of the field toroidal component occurs
only in the northern hemisphere, within the first two years after activity minimum. This asymmetry is
the result of the modulation range difference between solar hemispheres. 

Similar to the previous case the modulation range decreases in equatorial area, where
the three-time sign reversal effect vanishes. But in this case the magnetic field cyclic oscillation modulation range 
is maximal in the northern polar area. Hence, in this area modulation remains significant. Moreover,
near the northern pole the three-time sign reversal of the magnetic field meridional component becomes possible.

Further, we considered the excitement of the linear Rossby and Alfv\'en waves in the base of convection zone,
when in unperturbed state VSQBO and corresponding modulation of the magnetic field cyclic
oscillations are taken into account. We carried out the analytical and numerical investigation of the excitement and
evolution in time of these waves and we obtained the following results:

3) When in eqs. (\ref{f34})-(\ref{f35}) $k_{y1}=0$, i.e. $K_{21}=0$ in the environment
propagate only Rossby waves. In this case it is possible to analytically solve the set of
differential equations (\ref{f36})-(\ref{f39}) and we have obtained a solution of the meridional
component of the perturbed velocity (see eq. (\ref{f42})). By using this solution
we determined the expression of the Rossby wave hydrodynamical energy density temporal
variation.

4) We obtained the curves of the Rossby wave energy density time dependence. Curves in
Figure 3 correspond to the definition of the VSQBO amplitude by eq. (\ref{f22})
and Figure 4 corresponds to that defined by eq. (\ref{f24}). These figures show that the Rossby
wave energy density increases during very short time intervals and such behaviour of the
wave energy density are shown as sharp peaks. These peaks appear in the time moments when
the sign reversal of the unperturbed magnetic field toroidal component occurs. Thus, in time moments
when the three-time sign reversal of the magnetic field toroidal component happens we have three peaks
instead of one. In the case of $k_{y1}=0$ these time moments coincide with those of
the wavevector meridional component $K_y =0$. This means that in these moments Rossby
waves propagate exactly in the toroidal direction.

5) We performed the numerical investigation of the set of differential equations (\ref{f36})-(\ref{f39})
in case when $k_{y1}\neq 0$. In this case propagation of both Rossby and Alfv\'en waves is possible.
We obtained the curves of the wave hydrodynamic and magnetic energy density temporal variation. Figure 5
corresponds to the case "a" of the VSQBO amplitude and Figure 6 corresponds to case "b". These
results clearly show oscillation of the wave hydrodynamic and magnetic energy densities. By choosing the
$k_{x1}$ and $k_{y1}$ values one can obtain oscillation of wave hydrodynamical and magnetic energy
densities with periods comparable with those obtained from observations of the solar flare activity
(e.g see \opencite{bai87}; \opencite{ri84}). These reasons allow us to think that the model we have presented here
well describes the solar flare excitement and their periodical activity. Our results are in agreement with
the observations of solar flare activity which show the difference of the flare activity intensity between northern
and southern hemispheres of the Sun \cite{bai87}.

The model presented here is applicable for investigation of the magnetic field variations and the  linear wave
evolution in the layers where the arbitrary period oscillations of the velocity shear, characteristic to Solar atmosphere, occur.
In this sense it is important to study the role of the shear layer just below the photosphere in the distribution of 
the magnetic field at the solar surface. This model is also useful for the investigation of
the excitement and evolution of waves close to the solar surface and their influence on the solar corona.
Finally, taking into account the velocity shear oscillation with 11-year period, these work results could be used
for investigation of solar p-mode properties during activity cycles. These very important tasks
will be subjects of our forthcoming works.

\begin{acknowledgements} 
The work was supported in part by INTAS grant No 31931; B.M.Sh. addresses thanks to
D. Grilli for carefully reading of the manuscript and suggested corrections. 
\end{acknowledgements}

\end{article}

\begin{thebibliography}{}
\bibitem[\protect\citeauthoryear{Akioka et al.}{1987}]{ak87}
Akioka, M., Kubota, J., Ichimoto, K., Tohmura, I. and Suzuki, M.:1987, {\it Solar Phys.} {\bf 112}, 313
\bibitem[\protect\citeauthoryear{Ambroz}{1973}]{am73}
Ambroz, P.:1973, {\it Bull. Astron. Inst. Czechoslovakia} {\bf 24}, 80
\bibitem[\protect\citeauthoryear{Bai}{1995}]{bai95}
Bai, T.:1995, {\it in IAU Colloq. 153, Magnetodynamic Phenomena in the Solar Atmosphere:\\Prototypes of Stellar Magnetic
Activity}, ed. Y. Uchida (Dordrecht: Kluwer), 337
\bibitem[\protect\citeauthoryear{Bai}{1987}]{bai87}
Bai, T.:1987, {\it Astroph. J.}, {\bf 318}, L85
\bibitem[\protect\citeauthoryear{Benevolenskaya}{1998}]{be98}
Benevolenskaya, E. E.:1998, {\it Solar Phys.} {\bf 181}, 479
\bibitem[\protect\citeauthoryear{Benevolenskaya}{1996}]{be96}
Benevolenskaya, E. E.:1996, {\it Solar Phys.} {\bf 167}, 47
\bibitem[\protect\citeauthoryear{Benevolenskaya}{1995}]{be95}
Benevolenskaya, E. E.:1995, {\it Astron. Lett.} {\bf 21}, 495
\bibitem[\protect\citeauthoryear{Benevolenskaya}{1994}]{be94}
Benevolenskaya, E. E.:1994, {\it Astron. Lett.} {\bf 20 } (7), 551
\bibitem[\protect\citeauthoryear{Benevolenskaya et al.}{1999}]{bek99}
Benevolenskaya, E. E., Hoeksema, J. T., Kosovichev, A. G. and Scherrer, P. H.:\\1999, {\it Astrophys. J.} {\bf 517}, L163
\bibitem[\protect\citeauthoryear{Benevolenskaya and Makarov}{1992}]{be92}
Benevolenskaya, E. E. and Makarov, V. I.:1992, {\it Sov. Astr. Lett.} {\bf 18} (2), 108
\bibitem[\protect\citeauthoryear{Brown et al.}{1989}]{br89}
Brown, T. M., Christensen-Dalsgaard, J., Dziembowski, W. A., Goode, P., Gough, D. O. and Morrow, C.A.:\\1989,
{\it Astrophys. J.} {\bf 343}, 526
\bibitem[\protect\citeauthoryear{Bumba}{1989}]{bu89}
Bumba, V.:1989, {\it Bull. Astron. Inst. Czechoslovakia} {\bf 40}, 17
\bibitem[\protect\citeauthoryear{Chagelishvili et al.}{1993}]{ch93}
Chagelishvili, G. D., Christov, T.S., Chanishvili R.G. and Lominadze, J.G.:1993, {\it Phys. Rev (E)} {\bf 47}, 366
\bibitem[\protect\citeauthoryear{de Toma et al.}{2000}]{to20}
de Toma, G., White, O. R. and Harvey, K. L.:2000, {\it Astrophys. J.} {\bf 529}, 1101
\bibitem[\protect\citeauthoryear{Farrell}{1987}]{fa87}
Farrell, B.:1987, {\it J. Atmos. Sci.} {\bf 44}, 2191
\bibitem[\protect\citeauthoryear{Gaizauskas et al.}{1983}]{ga83}
Gaizauskas, V., Harvey, K. L., Harvey, J. W. and Zwaan, C.:1983, {\it Astrophys. J.} {\bf 265}, 1056
\bibitem[\protect\citeauthoryear{Gigolashvili et al.}{1995}]{gi95}
Gigolashvili, M. Sh., Japaridze, D. R., Pataraya, A. D. and Zaqarashvili T. V.:\\1995, {\it Solar Phys.} {\bf 156}, 221
\bibitem[\protect\citeauthoryear{Howard and Harvey}{1970}]{ho70}
Howard, R. and Harvey, J.:1970, {\it Solar Phys.} {\bf 12}, 23
\bibitem[\protect\citeauthoryear{Kosovichev and Schou}{1997}]{ko97}
Kosovichev, A. G. and Schou, J.:1997, {\it Astrophys. J. Lett.} {\bf 482}, L207
\bibitem[\protect\citeauthoryear{Lominadze et al.}{1988}]{lo88}
Lominadze, J. G., Chagelishvili, G. D. and Chanishvili, R.G.:\\1988, {\it Sov. Astr. Lett.} {\bf 14} (5), 364
\bibitem[\protect\citeauthoryear{Marcus and Press}{1977}]{mp77}
Marcus, P. and Press, W. H.:1977, {\it J. Fluid Mech.} {\bf 79}, 525
\bibitem[\protect\citeauthoryear{Parker}{1979}]{pa79}
Parker, E. N.:1979, {\it Cosmical Magnetic Fields}, Oxford University Press, England
\bibitem[\protect\citeauthoryear{Pataraya et al.}{1997}]{pa97}
Pataraya, A. D., Kaghashvili E. K and Pataraya, T. A.:\\1997,
{\it in JENAM-97,6$^{th}$ European and 3$^{rd}$ Hellenic Astronomical Conference}, abs. p. 50
\bibitem[\protect\citeauthoryear{Pataraya and Zaqarashvili}{1995}]{pz95}
Pataraya, A. D. and Zaqarashvili T. V.:1995{\it Solar Phys.} {\bf 157}, 31 
\bibitem[\protect\citeauthoryear{Priest}{1982}]{pr82}
Priest, E. R.:1982, {\it Solar Magnetohydrodynamics.}, by D. Reidel Publishing Company, Dordrecht, Holland
\bibitem[\protect\citeauthoryear{Rieger et al.}{1984}]{ri84}
Rieger, E., Share, G. H., Forrest, D. J., Kanbach, G., Reppin, C. and Chupp, E. L.:\\1984,
{\it Nature}, {\bf 312}, 623
\bibitem[\protect\citeauthoryear{Roberts}{1996}]{Ro96}
Roberts, B.:1996, {\it Bull. Astron. Soc. In.} {\bf 24}, 199
\bibitem[\protect\citeauthoryear{Sekii}{1998}]{se98}
Sekii, T.:1998, {\it Sounding Solar and Stellar Interiors}, Procc. 181$^{st}$ IAU sym.(Dordrecht: Kluwer), 189
\bibitem[\protect\citeauthoryear{Taylor}{1931}]{ty31}
Taylor, G. J.:1931, {\it Proc. R. Soc. London} {\bf A 132}, 499
\bibitem[\protect\citeauthoryear{Vitinskij}{1982}]{vi82}
Vitinskij, Y. I.:1982, {\it Soln. Dannye} {\bf 2}, 113
\end{thebibliography}
\end{document}